 \documentclass[aps,pra,notitlepage,floatfix,superscriptaddress,twocolumn]{revtex4-1}
\usepackage{amssymb, graphicx,subfigure}
\usepackage{enumerate,tensor}
\usepackage{amsmath,bm}
\usepackage{dcolumn}
\usepackage{color}
\usepackage{natbib} 
\usepackage[latin1]{inputenc}
\usepackage{subfigure}
\usepackage{textcomp}
\usepackage{hyperref}
\usepackage{booktabs}
\usepackage{makecell}
\usepackage{psfrag}
\usepackage{epsfig}
\usepackage[draft]{fixme}
   \fxsetup{layout=pdfcnote,mode=multiuser}
   \FXRegisterAuthor{rg}{arg}{RGF}

\usepackage[per-mode=symbol]{siunitx}   
\usepackage[version=3]{mhchem}
 
\DeclareSIUnit\intensity{\watt\per\centi\meter\squared}
\DeclareSIUnit\fieldstrength{\volt\per\centi\meter}
\DeclareSIUnit\kfieldstrength{k\volt\per\centi\meter}
\DeclareSIUnit\energy{cm^{-1}}

\newcommand{\degree}{\ensuremath{^\circ}}%
\newcommand{\eg}{e.\,g.}%
\newcommand{\Estat}{\ensuremath{\mathbf{\text{E}}_{s}}}%
\newcommand{\Estatabs}{\ensuremath{\text{E}_{s}}}%
\newcommand{\ie}{i.\,e.}%
\newcommand{\Inotime}{\textup{I}}%
\newcommand{\Ialign}{\textup{I}_{0}}%
\newcommand{\Nuptotal}{\ensuremath{\text{N}_\text{up}/\text{N}_\text{tot}}}%
\newcommand{\pstate}[4]{\ensuremath{\left|#1_{#2,#3}#4\right>}}%
\newcommand{\ppstate}[4]{\ensuremath{\left|#1_{#2,#3}#4\right>_\text{p}}}%
\newcommand{\ptstate}[4]{\ensuremath{\left|#1_{#2,#3}#4\right>_\text{t}}}%

\makeatletter
\def\paragraph{\@startsection{paragraph}{4}{10pt}{-1.25ex plus -1ex minus -.1ex}{0ex plus 0ex}{\normalsize\textit}}
\renewcommand\@biblabel[1]{#1}
\renewcommand\@makefntext[1]%
{\noindent\makebox[0pt][r]{\@thefnmark\,}#1}
\DeclareRobustCommand\onlinecite{\@onlinecite}
\def\@onlinecite#1{\begingroup\let\@cite\NAT@citenum\citealp{#1}\endgroup}
\def\tagform@#1{\maketag@@@{\ignorespaces#1\unskip\@@italiccorr}}
\let\orgtheequation\theequation
\def\theequation{(\orgtheequation)}
\makeatother

\newcommand{\ket}[1]{\left|#1\right\rangle}

\newcommand{\expectation}[3]{\left\langle #1\left|#2\right|#3\right\rangle}
\newcommand{\expected}[1]{\left\langle #1\right\rangle}
\newcommand{\granada}{\affiliation{Instituto Carlos I de F\'{\i}sica Te\'orica y Computacional and
      Departamento de F\'{\i}sica At\'omica, Molecular y Nuclear, Universidad de Granada, 18001
      Granada, Spain}}%
\newcommand{\aarphys}{\affiliation{Department of Physics and Astronomy, Aarhus University, 8000
      Aarhus C, Denmark}}

\begin{document}

\title{Theoretical description of mixed-field
orientation of asymmetric top molecules: a time-dependent study}

\author{Juan J.\ Omiste}\aarphys
\author{Rosario Gonz\'alez-F\'erez}\granada

\date{\today}
\begin{abstract} 
We present a theoretical study of the 
mixed-field-orientation  of asymmetric top molecules
in tilted  static electric field and non-resonant  linearly polarized laser pulse  by
solving the time-dependent Schr\"odinger equation.
Within this framework, we compute the mixed-field orientation of  a state selected molecular 
beam of  benzonitrile (C$_7$H$_5$N) and 
compare with the experimental observations~\cite{PhysRevA.83.023406}, 
and with our previous time-independent descriptions~\cite{omiste:pccp2011}. 
For an excited rotational state,
we investigate the field-dressed dynamics for several field configurations as
those used in the mixed-field experiments.  The non-adiabatic phenomena and their consequences on 
the rotational dynamics  are analyzed in detail. 
\end{abstract}
\maketitle
\section{Introduction}
\label{sec:introduction}

During the last years, experimental efforts have been undertaken to 
develop and improve experimental techniques to  enhance
the orientation of polar molecules~\cite{kupper:prl102,nevo:pccp11,frumker2012,Trippel:MP111:1738,PhysRevA.89.051402,Kraus_jpb_2014}.
When a molecule is oriented 
the molecular fixed axes are confined  along the laboratory fixed axes
and its permanent dipole moment possesses a well defined direction.
The  experimental efforts are motivated by
the broad range of promising perspectives and possible applications
of   oriented  molecules, such as 
high-order harmonic generation~\cite{Peng2013,Zhang2015,Li2016a},
chemical reaction  dynamics~\cite{loesch:9016,Rakitzis:Science303:1852,Ospelkaus853,Quemener2010a}, 
ultracold molecule-molecule collision dynamics~\cite{loesch:jcp93,Ni2010,Henson234,Vogels787},
and  diffractive  imaging of polyatomic molecules~\cite{Yamazaki2015,Kierspel2015a}.

The theoretical prediction to strongly orient molecules by coupling the quasidegenerate levels of a non-resonant-laser generated pendular state~\cite{Friedrich1999,friedrich:jcp111}, quickly became a promising experimental 
technique~\cite{baumfalk:jcp114-a,sakai:prl_90,PhysRevA.72.063401}.
However, only using state-selected ensembles of linear and asymmetric top 
molecules unprecedented degrees of orientation could be reached~\cite{kupper:prl102,kupper:jcp131,nevo:pccp11,ghafur_impulsive_2009}. 
These experimental efforts have been accompanied of theoretical studies to provide a better 
physical insight into the field-dressed dynamics. 
For a state-selected beam of
asymmetric top molecules, the first analysis 
 showed that the 
experimental mixed-field orientation could not be reproduced in an adiabatic description~\cite{omiste:pccp2011}.
Based on the lack of  azimuthal symmetry due to the 
 weak static electric field, a diabatic model was proposed to classify 
 the avoided crossing as diabatic and adiabatic depending on 
 the field-free magnetic quantum numbers of the involved states~\cite{omiste:pccp2011}.
An explicit  time-dependent analysis of the mixed-field-orientation experiments of OCS
concluded that this process is, in general, non-adiabatic and requires a time-dependent 
quantum-mechanical description~\cite{nielsen:prl2012}.
The lack of adiabaticity is due to the formation of the quasidegenate pendular doublets
as the laser intensity is increased, the resulting  
narrow avoided crossings, and  the corresponding couplings between the states in a $|J,M\rangle$ manifold 
for tilted fields~\cite{nielsen:prl2012,omiste:pra2012}.  
These non-adiabatic phenomena provoke a transfer of population between 
energetically neighboring  adiabatic pendular states, which might significantly reduce the degree of 
orientation~\cite{nielsen:prl2012,omiste:pra2012},
 this effect can be mitigated  using stronger dc electric fields~\cite{CPHC:CPHC201600710}. 
This population transfer between the oriented and anti-oriented states  forming a pendular 
doublet could be efficiently controlled to achieve a strong
field-free orientation during the post-pulse dynamics~\cite{PhysRevLett.114.103003}.

For an asymmetric top molecule, we had performed a first time-dependent study on parallel
 fields showing the complexity of the field-dressed dynamics~\cite{omiste:pra_88_2013}. 
Here, we extend this work  and  analyze the rotational dynamics of benzonitrile (BN) in  
tilted-field configurations similar to those used in current mixed-field experiments.
 Within this time-dependent framework, we revisit  the experiment on mixed-field orientation of 
 a state-selected molecular beam of benzonitrile~\cite{PhysRevA.83.023406,Holmegaard:natphys6},
extending upon our earlier theoretical description~\cite{omiste:pccp2011}.
The sources of discrepancies between the experimental results and the
time-dependent description are analyzed. 
For a prototypical excited rotational state, we explore  the field-dressed rotational dynamics and investigate
the complicated  non-adiabatic phenomena, showing that it is experimentally more challenging 
 to reach the adiabatic limit.

The paper is organized as follows: In \autoref{sec:hamiltonian} we present the
Hamiltonian of a polar asymmetric top molecule in tilted electric and non-resonant laser fields,
 and the numerical method used to solve the time-independent Schr\"odinger
equation. 
The time-dependent description of the mixed-field orientation experiment of a
benzonitrile molecular beam
is presented in~\autoref{sec:orientation_molecular_beam}, where we also provide a 
comparison with our previous time-independent analysis. 
Section~\ref{sec:field_dressed_dynamics} is devoted to investigate the mixed-field dynamics of 
an  excited rotational state, and we explore the sources of non-adiabatic effects in~\autoref{sec:sources}. The conclusions are given in~\autoref{sec:conclusions}.

\section{The Hamiltonian of an asymmetric top molecule in tilted fields}
\label{sec:hamiltonian}

We consider a polar asymmetric top molecule in combined electric and non-resonant laser 
fields.
Our study is restricted to  asymmetric top molecules that can be described within the rigid rotor 
approximation, and that have the polarizability tensor diagonal in the principal axes of inertia frame 
and  the permanent electric  
dipole moment  $\mathbf{\mu}$  parallel to  the $z$-axis of this molecule fixed frame (MFF) $(x,y,z)$. 
The MFF  $z$-axis is taken along the molecular axis with  the smallest moment of  inertia.
The non-resonant laser field is linearly polarized along the $Z$-axis of the
laboratory fixed frame (LFF) $(X,Y,Z)$ and the electric field is tilted by an angle $\beta$ 
with respect  to the LFF $Z$-axis 
and  contained in the $XZ$-plane.
The rotational dynamics of the molecule is described 
using the Euler angles $(\theta, \phi,\chi)$ that relate the LFF  
and MFF~\cite{zare}. In the framework of the rigid rotor approximation, 
the Hamiltonian reads as
\begin{equation}
  \label{eq:hamiltonian_total}
  H(t)=H_R+H_s(t)+H_L(t),
\end{equation}
where $H_R$ is the field-free rotational Hamiltonian
\begin{equation}
H_R=B_x{J}_x^2+B_y{J}_y^2+B_z{J}_z^2 
  \label{eq:hamiltonian_rot}
\end{equation}
with ${J}_k$ being the projection of the total angular momentum operator $\mathbf{J}$
along the  MFF $k$-axis with $k=x,y$, and $z$,
and $B_k$
the rotational constant along the MFF $k$-axis.
For  BN, the rotational
constants are $B_{x}=1214$~MHz, $B_{y}=1547$~MHz and 
$B_{z}=5655$~MH~\cite{wohlfart:jms247}. 

The electric field $\Estat(t)$ interacts with the electric dipole moment $\mathbf{\mu}$ of the  molecule as
\begin{equation}
  \label{eq:hamiltonian_stark}
  H_s(t)=-\mu \Estatabs(t)\cos\theta_s, 
\end{equation}
where $\theta_s$ is the angle  between the electric field and the MFF $z$-axis, 
$\cos\theta_s=\cos\theta\cos\beta+\sin\beta\sin\theta\cos\phi$.
For  BN, the  permanent  electric 
dipole moment is  $\mu=4.515$~D~\cite{wohlfart:jms247}.
 $\Estatabs(t)$ initially depends linearly on time, and once the maximum strength
$ \Estatabs$ is reached, it  is kept constant. 
The turning on speed ensures that the process is
adiabatic, and we have neglected
 the coupling of this field with the molecular polarizability or higher order terms.

For a non-resonant linearly polarized laser field, 
 the interaction reads~\cite{seideman2006,omiste:jcp2011}
\begin{equation}
  \label{eq:hamiltonian_laser}
 H_L(t)=-\cfrac{\textup{I}(t)}{2\epsilon_0c}
 \left(\alpha^{zx}\cos^2\theta+\alpha^{yx}\sin^2\theta\sin^2\chi\right),
\end{equation}
where $\alpha^{km}=\alpha_{kk}-\alpha_{mm}$ are the polarizability
anisotropies, and $\alpha_{kk}$ the polarizability along the molecular $k$-axis $k,m=x, y$ and $z$.
For  BN,  the polarizabilities are $\alpha_{xx}=7.49$~\AA$^3$, 
$\alpha_{yy}=13.01$~\AA$^3$ and 
$\alpha_{zz}=18.64$~\AA$^3$~\cite{wohlfart:jms247}. 
$\epsilon_0$ is the dielectric constant and $c$ the speed of light.
In this work, we  consider Gaussian pulses with intensity
 $\textup{I}(t) = \Ialign\exp\left(-\cfrac{4\ln 2 t^2}{\tau^2}\right)$, 
$\Ialign$ is the peak intensity and $\tau$ is the full width half maximum (FWHM).

For  tilted fields with $\beta\ne0\degree,\,90\degree,\,180\degree$, 
the symmetries of the rigid rotor  Hamiltonian~\eqref{eq:hamiltonian_total}
are the identity, $E$, the two fold rotation around the MFF $z$-axis, $C_z^2$, and the reflections $\sigma_{XZ}$  on 
the plane spanned by the two fields, \ie, the $XZ$-plane. These symmetries imply the conservation of the parity of the 
projection of  $\mathbf{J}$ on the MFF $z$-axis, \ie, the parity of $K$, and the
parity under the reflections on the LFF $XZ$-plane $\sigma_{XZ}$. Consequently, the eigenstates can be classified in
four different irreducible representations~\cite{omiste:jcp2011},
whose basis elements are presented in~\autoref{tb:irreps} 
 in terms of the field-free symmetric top eigenstates $|JKM\rangle$~\cite{omiste:jcp2011}.
\begin{table}
  \begin{ruledtabular}
\begin{tabular}{lllc}
 \multicolumn{3}{c}{Parity}&  Functions\\
    $C_z^2$ & $\sigma_{XZ}$ & $K$ & \\
\hline
$e$ & $e$ &  $e$, $K,M=0$ & $\ket{J00}$\\
\hline
$e$ & $e$ & $e$ & $\frac{1}{\sqrt{2}}\left(|JKM\rangle +
  (-1)^{K+M}|J-K-M\rangle\right) $\\ 
\hline
$e$ & $o$ & $e$ & $\frac{1}{\sqrt{2}}\left(|JKM\rangle + (-1)^{K+M+1}|J-K-M\rangle\right)$\\
\hline
$o$ & $e$ & $o$& $\frac{1}{\sqrt{2}}\left(|JKM\rangle + (-1)^{K+M}|J-K-M\rangle\right)$\\
\hline
$o$ & $o$ & $o$& $\frac{1}{\sqrt{2}}\left(|JKM\rangle + (-1)^{K+M+1}|J-K-M\rangle\right)$\\
\end{tabular}
  \end{ruledtabular}
\caption{For tilted fields with $\beta\ne 90\degree$, functions used in the 
basis set expansion of the time-dependent wave function. 
\label{tb:irreps} }
\end{table}

The time-dependent Schr\"odinger equation associated with the Hamiltonian~\ref{eq:hamiltonian_total}
 is solved combining the short iterative Lanczos  method~\cite{Beck:phys_rep_324} for the time variable and 
a basis set expansion in the field-free top eigenstates written in terms of the Wang
states, $\ket{JKM}$, for the angular coordinates.
For each irreducible representation, the  symmetric top wave functions forming the basis are 
properly symmetrized as indicated  in~\autoref{tb:irreps}~\cite{omiste:jcp2011}. 
The time-dependent wave function is labeled using the adiabatic following and the field-free 
notation  $\ptstate{J}{K_a}{K_c}{M}$
where $ K_a$ and $K_c$ are the values of $K$ for the limiting prolate and oblate
symmetric top rotor,  respectively~\cite{king_jcp11}.

To analyze the rotational dynamics,  the time-dependent 
wave function is projected on the adiabatic pendular states  at time $t$
\begin{equation}
  \label{eq:projection_wf}
 \ptstate{J}{K_a}{K_c}{M} =\sum 
 C_{\pstate{{\cal J}}{{\cal K}_a}{{\cal K}_c}{\cal{M}}}(t)
\ppstate{{\cal J}}{{\cal K}_a}{{\cal K}_c}{\cal{M}}
 \end{equation}
with  $C_{\pstate{{\cal J}}{{\cal K}_a}{{\cal K}_c}{\cal{M}}}(t)=
{}_\textup{p}\left<\right.{\cal J}_{{\cal K}_a{\cal K}_c}{\cal{M}}\ptstate{J}{K_a}{K_c}{M}$
and  the wave function of the adiabatic
pendular state of the instantaneous Hamiltonian~\eqref{eq:hamiltonian_total}
\ppstate{{\cal J}}{{\cal K}_a}{{\cal K}_c}{\cal{M}} that connects 
adiabatically to the field-free state \pstate{{\cal J}}{{\cal K}_a}{{\cal K}_c}{\cal{M}}. 
The sum in~\autoref{eq:projection_wf} runs over all  pendular states 
 within the same irreducible representation.

The  rotational dynamics can be characterized by 
 the adiabaticity ratio or parameter~\cite{ballentine:quantum_mechanics} 
\begin{equation}
\label{eq:eta}
\eta=\cfrac{\hbar\left|\tensor[_{\textup{p}}]{\left\langle k \left|\cfrac{\partial H_\textup{L}(t)}{\partial t}\right| m\right\rangle}{_{\textup{p}}}
\right|}{\left|E_m-E_k\right|^2}
\end{equation}
where $\left|k\right\rangle_\text{p}$ and $\left|m\right\rangle_\text{p}$ are the 
eigenfunctions of the  adiabatic pendular eigenstates of 
Hamiltonian~\ref{eq:hamiltonian_total} at time $t$ and $E_k$ and $E_m$ are the eigenenergies. 
The rotational dynamics can be considered as adiabatic at a certain time if 
$\eta\ll 1$~\cite{ballentine:quantum_mechanics}.

\subsection{Experimental measures}
\label{subsec:experiment}

In  mixed-field orientation experiments, the degree of orientation can be  measured by a
multiple-ionization and a subsequent  Coulomb explosion  of the molecule, and 
the velocity mapping  of the ionic fragments onto a 2D screen perpendicular to the dc field of the 
velocity mapping electrodes. 
The molecular orientation is reflected in the 2D-images, which show  an
up/down asymmetry measured  by the ratio
$\Nuptotal$, where $\text{N}_\text{tot}$ and $\text{N}_\text{up}$  stand for the amount of ions 
collected on the full screen and  on its upper part, 
respectively, \cite{kupper:prl102,kupper:jcp131,nevo:pccp11}. 
Theoretically, the recorded  image corresponds to the 2D projection  of
the 3D probability density on the 
 screen perpendicular to the  electric field~\cite{omiste:pccp2011,nielsen:prl2012}.
The theoretical orientation ratio $\Nuptotal$ reads
\begin{equation}
\label{eq:nupntot_theory}
\Nuptotal=\cfrac{\int_0^\infty\int_{-\infty}^\infty\rho(y_s,z_s)dy_sdz_s}{\int_{-\infty}^\infty\int_{-\infty}^\infty\rho(y_s,z_s)dy_sdz_s}.
\end{equation}
where  $\rho(y_s,z_s)$ is the 2D probability density, $y_s$ and $z_s$ are the horizontal and vertical screen coordinates, respectively.
Here, we derive  this 2D probability density including the  selectivity factor of a circularly 
polarized probe laser and the experimental velocity distribution of the ionic fragments under the 
recoil approximation~\cite{omiste:pccp2011}. Note that for non-oriented states, $\left\langle\cos\theta\right\rangle=0$ 
and $\Nuptotal=0.5$, whereas $\Nuptotal>0.5$ and $\Nuptotal<0.5$ for oriented and antioriented states, respectively.

\section{Mixed-field orientation for experimental conditions} 
\label{sec:orientation_molecular_beam}

In this section we present a theoretical time-dependent description of the  mixed-field 
orientation corresponding to previous mixed-field-orientation experiments of benzonitrile~\cite{kupper:jcp131}, and
a comparison with our previous time-independent analysis~\cite{omiste:pccp2011}. 
State-selected benzonitrile molecules in a molecular beam were
  oriented using the weak dc  electric  field from the velocity-map-imaging spectrometer and
a non-resonant laser pulse~\cite{kupper:jcp131}. 
For a $10$~ns  laser pulse with peak intensity 
$\Ialign=\SI{7e11}{\intensity}$ and a weak dc field, 
$\Estatabs=\SI{286}{\fieldstrength}$,  tilted  with respect to each other an angle   $\beta=135\degree$, 
 the experimentally measured degree of orientation was
$\Nuptotal=0.71$~\cite{PhysRevA.83.023406}.
For this field configuration, we perform a theoretical study  including $54$  rotational states of
the  state-selected molecular beam of this experiment,
\ie, accounting for $92.5\%$ of the total population in the  beam~\cite{PhysRevA.83.023406}.

The adiabatic approximation, which assumes that the field-dressed instantaneous eigenstate is a solution of the time-dependent Hamiltonian,
 would give rise to a very weakly oriented ensemble with $\Nuptotal=0.55$~\cite{omiste:pccp2011}.
This result is in contradiction with
the experimental observation and indicates that the mixed-field orientation is, in general, a non 
adiabatic process~\cite{omiste:pccp2011}. 
The weak electric field is  responsible for breaking the azimuthal symmetry, and for  
coupling  states with different field-free magnetic quantum numbers.
Based on this fact, we had proposed a diabatic model~\cite{omiste:pccp2011}, 
which  improves the adiabatic description by classifying the 
avoided crossings: as $\Inotime(t)$ increases, 
they are crossed diabatically (adiabatically) if the involved
states have different (same) field-free expectation value $\expected{M^2}$.
The diabatic model is equivalent to an adiabatic description of a parallel field configuration 
including  only the dc electric field  component  parallel to the alignment laser field
$\Estatabs\cos\beta$, and neglecting  the perpendicular component. 
Within this diabatic model, 
the orientation ratio of the state-selected molecular beam   is 
$\Nuptotal=0.624$, which is larger than for the pure adiabatic description, but still   smaller than the experimental measurement.

To check the validity of the diabatic model, we have solved
the time-dependent Schr\"odinger equation considering a dc field   
parallel to the LFF $Z$-axis with strength $\Estatabs\cos\beta$.
Hence, the field-dressed rotational dynamics takes into account 
the non-adiabatic couplings between states with the same magnetic
quantum number and  the  pendular doublets formation~\cite{omiste:pra_88_2013}. 
In this  description, we obtain a smaller  orientation ratio  $\Nuptotal=0.623$
than the experimentally measured. 
For a  full non-adiabatic description,
we  have taken into account all the couplings, and, in particular, those 
between states with different field-free magnetic quantum numbers,
 which were neglected in the parallel-field  time-dependent  description,  
and solved the time-dependent Schr\"odinger equation associated to 
Hamiltonian \eqref{eq:hamiltonian_total} for each initially populated state. 
The obtained degree of orientation  $\Nuptotal=0.612$ is still lower than the experimental one,
and even lower than the simplified models described above.
The discrepancies between the time-dependent descriptions of the tilted and parallel fields 
configurations  arise  
due to the couplings between  the states within the \pstate{J}{K_a}{K_c}{M}-manifold with
$0\le M\le J$, \ie, degenerate states
in the field-free case, which differ in the magnetic quantum number $M$, at
weak laser intensities, \ie, $I\lesssim \SI{1e18}{\intensity}$, and between the states with
 different field-free magnetic quantum numbers at strong  intensities~\cite{omiste:pra2012}.

\begin{figure}
 \includegraphics[width=0.95\linewidth]{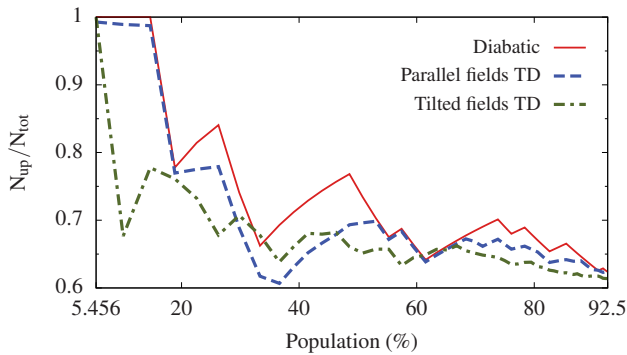}
\caption{The theoretical orientation ratio $\Nuptotal$ as a function of the population
  of the molecular beam of benzonitrile using a probe laser circularly polarized perpendicular to the screen   computed by
   the diabatic model (red solid line), 
  by time-dependent (TD) description  for parallel-fields
 (blue dashed line) and time-dependent description  for tilted-fields 
  (green dot-dashed line).
 The field configuration is 
  $\Ialign=\SI{7e11}{\intensity}$, $\Estatabs=\SI{286}{\fieldstrength}$ and $\beta=135\degree$.}
 \label{fig:nup_ntot_cr}
 \end{figure}

To analyze these theoretical results, 
we build up the molecular ensemble by successively adding states
according to their populations in the experimental state-selected beam. The  orientation ratio  $\Nuptotal$ is plotted in Fig.~\ref{fig:nup_ntot_cr}
versus the percentage of population included in the experimental molecular beam. 
For the ground state, which  has the largest population, we obtain $\expected{\cos\theta}=0.986$ and 
$\Nuptotal=0.999$, which is very close to the adiabatic result  $\Nuptotal=1$.
As more states are included in the ensemble, the orientation ratio $\Nuptotal$ decreases  with a superimposed oscillatory 
behavior due to the orientation or  antiorientation of the additionally included states. 
The discrepancies between these results illustrate the importance of performing a full  time-dependent description of the
mixed field orientation process. 

Several reasons could explain the disagreement between the time-dependent study and 
the experimental result.
 First, we do not include the finite spatial profile of the alignment and probe lasers which implies 
 that all the  molecules in the beam do not feel the same laser intensity. Based on our previous 
 studies~\cite{omiste:pccp2011},  we can conclude that this effect should not significantly modify the degree of 
 orientation.
  Second, it has been assumed that the state selection  along the dc-field deflector
is an adiabatic process~\cite{kupper:prl102,kupper:jcp131,nevo:pccp11}, if this would not be the case, the rotational states before the mixed-field
orientation  experiment would not be pure field-free states.
Third, the mixed-field dynamics is very sensitive to the field configuration, and small variations 
on it, \eg,  on the pulse shape,  could significantly affect these results. 
Finally, by adding the rest of states forming  the experimental molecular
beam, the theoretical degree of orientation will not 
approach to the  experimental one. 
Assuming that the rest of the states are not oriented, or that half of them are fully oriented and
the other half fully antioriented, the theoretical degree of orientation would be reduced to $0.605$, still far from
the experimental value.

 \begin{figure}
 \includegraphics[width=0.95\linewidth,angle=0]{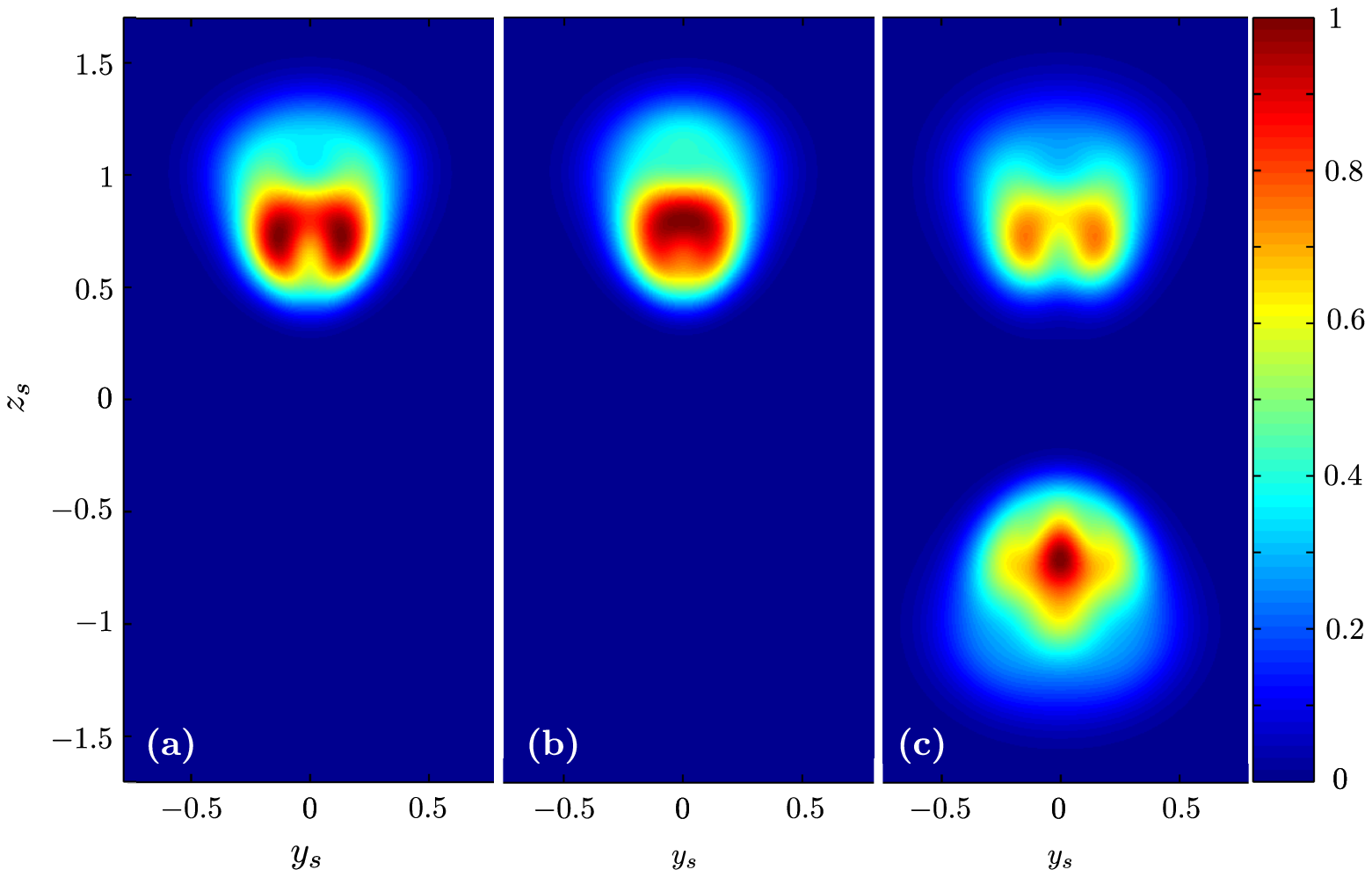}
\caption{The 2D projection of the probability density of the state $\ptstate{3}{0}{3}{1}$ 
at the peak intensity computed using (a) the adiabatic approximation, (b) the diabatic model, and (c) the time-dependent description for tilted fields. The field configuration is 
$\Ialign=\SI{7e11}{\intensity}$, $\Estatabs=\SI{286}{\fieldstrength}$ and $\beta=135\degree$.
}
 \label{fig:screen_03_00_03_01}
 \end{figure}
The adiabatic and diabatic approximations predict 
pendular states either fully oriented $\Nuptotal\approx 1$ or antioriented 
$\Nuptotal\approx 0$, 
 whereas in a time-dependent description  they are not fully oriented or fully antioriented with
 $0\lesssim\Nuptotal\lesssim1$. 
 This is illustrated 
in~\autoref{fig:screen_03_00_03_01} with  the 2D projection of the probability density
of the $\ptstate{3}{0}{3}{1}$ state   at the peak intensity. 
 For the adiabatic and diabatic  descriptions, 
  the 2D probability density is concentrated on the upper part of the screen
 with $\Nuptotal=1$. 
In contrast,  the time-dependent description provides a weakly antioriented state with 
$\Nuptotal\approx 0.44$, see~\autoref{fig:screen_03_00_03_01}~(c). 
This is due to the contributions of antioriented pendular states to the 
rotational dynamics:
at the peak intensity  the antioriented adiabatic pendular state $\ppstate{2}{2}{0}{1}$ has
the largest contribution with a weight of $41.1\%$
into the  $\ptstate{3}{0}{3}{1}$ time-dependent wave function.

\section{Field-dressed dynamics of a rotational excited state}
\label{sec:field_dressed_dynamics}

 The rotational dynamics in tilted fields
 is more complex than in parallel fields~\cite{omiste:pra_88_2013}. 
We illustrate this complexity by  analyzing 
 the field-dressed dynamics of the excited state 
$\pstate{3}{0}{3}{1}$,  which has been studied for parallel fields in Ref.~\cite{omiste:pra_88_2013}. 
For tilted fields, the adiabatic pendular state 
$\ppstate{3}{0}{3}{1}$ is the ninth  one of the even-even  irreducible representation,
and  is oriented in a strong laser field combined with a weak static electric field.

 \begin{figure}
 \includegraphics[width=0.95\linewidth]{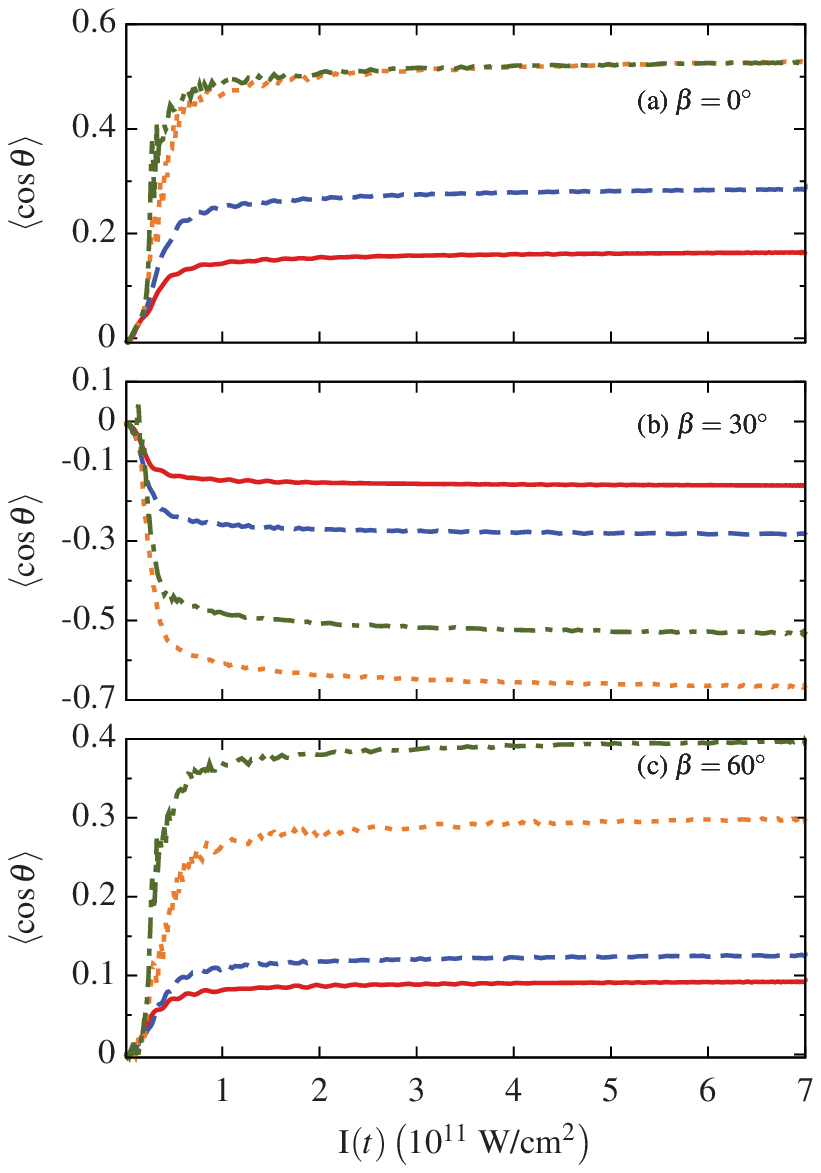}
 \caption{For the $\ptstate{3}{0}{3}{1}$ state, expectation value 
 $\left\langle\cos\theta\right\rangle$  as a function of $\textup{I}(t)$ for laser pulses with peak intensity 
 $\Ialign=\SI{7e11}{\intensity}$ and temporal widths $\tau=0.5$~(red solid), $1$~(blue dashed), $5$~(orange dotted) and $10$~ns~(green dot-dashed). The electric field has strength
  $\Estatabs=\SI{300}{\fieldstrength}$, and  is tilted an angle  (a)  $\beta=0\degree$,
   (b)  $\beta=30\degree$ and   (c)  $\beta=60\degree$.}
 \label{fig:j_3_03_1_es_300}
 \end{figure}

In Fig.~\ref{fig:j_3_03_1_es_300} 
we plot the orientation $\expected{\cos\theta}$ of the state $\ptstate{3}{0}{3}{1}$ as a function 
of the laser intensity $\Inotime(t)$ for 
three different field configurations (a) $\beta=0\degree$, (b) $\beta=30\degree$ and (c) 
$\beta=60\degree$,
with $\Estatabs=\SI{300}{\fieldstrength}$, and laser  pulses with temporal widths 
$\tau=0.5$~ns, $1$~ns, $5$~ns and $10$~ns and peak intensity  $\Ialign=\SI{7e11}{\intensity}$. 
For these field configurations,
the adiabatic orientation is given by  $\expected{\cos\theta}=0.97$, and 
weakly affected by the inclination angle of the dc-field $\beta$.

For parallel fields, $\ppstate{3}{0}{3}{1}$  is the  third adiabatic state in the irreducible  
representation with $M=1$ and even parity under two-fold  rotations around the MFF $z$-axis. 
This state is oriented in the pendular regime.   
As $\Inotime(t)$  increases, the orientation shows an increasing trend with a superimposed 
smooth oscillatory behavior, which is due to the couplings among the  adiabatic pendular states 
contributing  to the non-adiabatic  dynamics. Three adiabatic pendular states,  
$\ppstate{3}{0}{3}{1}$ (oriented), $\ppstate{2}{2}{1}{1}$ (antioriented), and 
$\ppstate{2}{2}{0}{1}$ 
(oriented), dominantly contribute to the time-dependent  wave function of  $\ptstate{3}{0}{3}{1}$.
The coupling between the oriented adiabatic pendular states, \ie, $\ppstate{3}{0}{3}{1}$ and 
$\ppstate{2}{2}{0}{1}$, provokes the oscillations, because in the pendular regime 
their couplings with the antioriented one are close to zero, \ie,  
${}_\textup{p}\expectation{3_{0,3}1}{\cos\theta}{2_{2,2}1}_\textup{p}\approx0$ and  
${}_\textup{p}\expectation{2_{2,0}1}{\cos\theta}{2_{2,1}1}_\textup{p}\approx0$~\cite{omiste:pra_88_2013}. 
The contribution of the antioriented adiabatic pendular state reduces the  degree of 
orientation compared to the adiabatic prediction. As the temporal width of the pulse $\tau$ 
increases, the dynamics becomes more adiabatic and  the orientation increases approaching to 
the adiabatic limit.

Due to the splitting of the field-free degenerate 
$\pstate{3}{0}{3}{M}$-multiplet in tilted fields, the adiabatic pendular states $
\ppstate{3}{0}{3}{M}$,
with $M=0,1, 2, 3$, contribute to the field-dressed dynamics of $\ptstate{3}{0}{3}{1}$ 
and their weights  depend  on the angle $\beta$ and on the dc-field  strength.
For $\beta=30\degree$, the $\ptstate{3}{0}{3}{1}$ wave function is antioriented, 
which can be rationalized in terms of the adiabatic pendular states contributing to the 
rotational dynamics. Once the multiplet  is split, the contribution of the 
adiabatic pendular states $\ppstate{3}{0}{3}{M}$  
onto the $\ptstate{3}{0}{3}{1}$ wave function  is   approximately the same for all the pulses. 
From this moment on, the way the laser is turned on plays an important role  on the dynamics. 
We find up to  $27$ adiabatic pendular states with $J\le 4$
contributing to the dynamics of $\ptstate{3}{0}{3}{1}$, 
and the antioriented adiabatic pendular state $\ppstate{2}{2}{0}{1}$ has the largest  contribution to the time-dependent 
wave function, 
larger than $67\%$ for $\tau=5$~ns and $10$~ns.  By decreasing the temporal width of the  pulse, more adiabatic pendular
states contribute to the  dynamics, and  $|\expected{\cos\theta}|$ diminishes
because  the weights of oriented and  antioriented adiabatic states are vey similar.
In contrast, $\ptstate{3}{0}{3}{1}$ is oriented if the dc-field is tilted an angle $\beta=60\degree$.
For this case, the dominant contributions to the wave function are due to the
  $\ppstate{2}{2}{1}{2}$ and $\ppstate{3}{2}{1}{3}$ oriented states for
  $\tau=10$~ns and $\tau=5$~ns, and to    $\ppstate{3}{2}{1}{3}$ for $\tau=1$~ns and 
  $\tau=0.5$~ns. As $\tau$ increases, the orientation increases, but it is  smaller than
the adiabatic limit of the orientation  $\expected{\cos\theta}=0.97$.

\begin{figure}
 \includegraphics[width=0.95\linewidth]{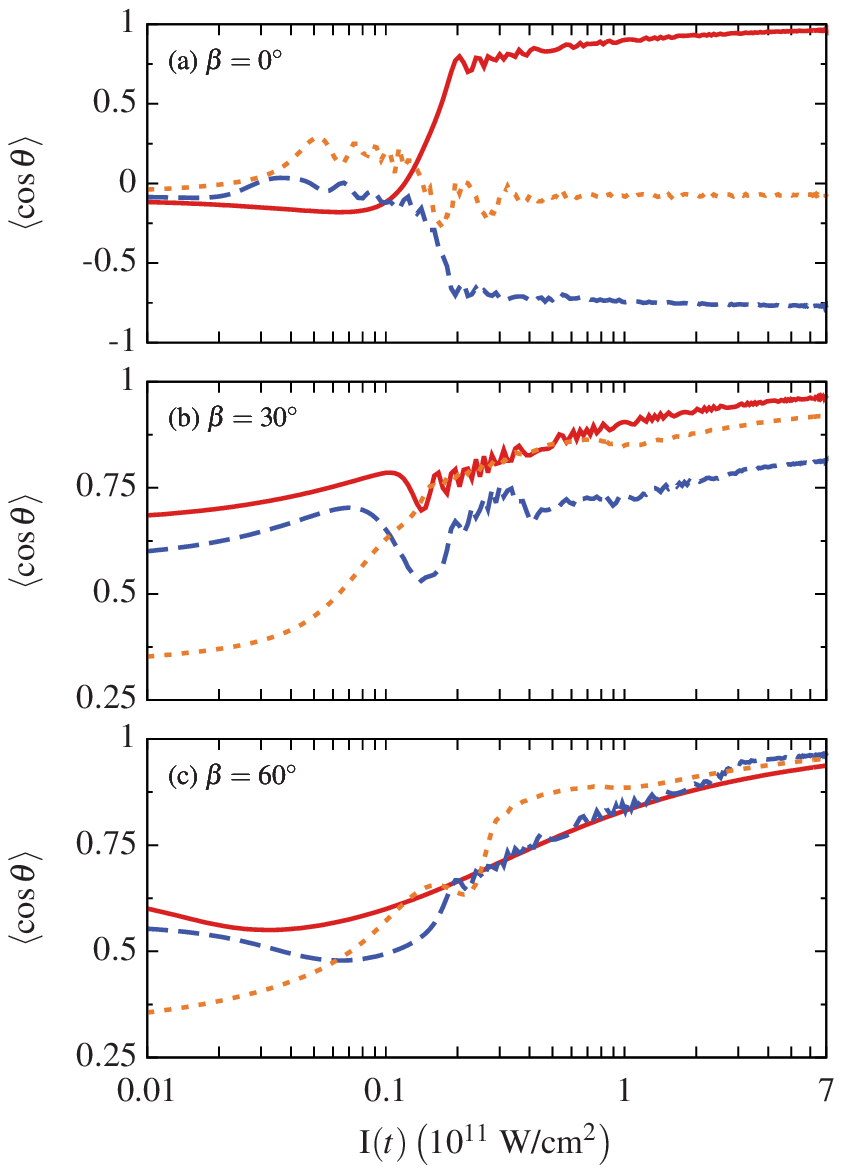}
 \caption{For the $\ptstate{3}{0}{3}{1}$ state, expectation value 
 $\left\langle\cos\theta\right\rangle$  as a function of $\textup{I}(t)$ for a $10$~ns
 laser pulse with peak intensity 
 $\Ialign=\SI{7e11}{\intensity}$. The electric field has strengths
 (a) $\Estatabs=\SI{5}{\kfieldstrength}$, 
 (b) $\Estatabs=\SI{10}{\kfieldstrength}$, and 
 (c) $\Estatabs=\SI{20}{\kfieldstrength}$, and 
 is  tilted by 
 $\beta=0\degree$ (red solid),  
 $\beta=30\degree$ (blue dashed) and  
 $\beta=60\degree$ (orange dotted).}
 \label{fig:j_3_03_1_dc_varios}
 \end{figure}
The dc-field strength determines the energy gap between the two states forming a pendular 
doublet in the strong laser field regime.
Thus, by increasing the dc-field strength the dynamics should become more adiabatic
 because the population transferred between 
these two states as the pendular doublet is formed should be reduced.
For a $10$~ns alignment pulse, we present in~\autoref{fig:j_3_03_1_dc_varios} 
the orientation of the $\ptstate{3}{0}{3}{1}$ state versus the laser intensity $\Inotime(t)$ 
for  dc-fields strengths
$\Estatabs=\SI{5}{\kfieldstrength}$, $\Estatabs=\SI{10}{\kfieldstrength}$ and 
$\Estatabs=\SI{20}{\kfieldstrength}$. 
Let us remark that for strong dc fields,
the energy gap between  two states forming the pendular doublets in a strong ac field is large and they are not 
quasidegenerate, 
and, in particular, the interaction due to the static electric field could not be treated as a perturbation to the ac-field interaction.
Even for these strong electric fields, the rotational dynamics is non-adiabatic,
and this state is either weakly oriented,  strongly oriented or antioriented depending on the field configuration, 
see~\autoref{fig:j_3_03_1_dc_varios}. 
This can be explained in terms the splitting of the $\ppstate{3}{0}{3}{M}$  manifold
and of the avoided crossings that are encountered during the 
time evolution of the wave packet, which, in most cases, are not crossed adiabatically~\cite{omiste:pra_88_2013}.
For strong dc fields, the avoided crossings between adiabatic pendular 
states evolving  from different multiplets become more likely.
For $\Estatabs=\SI{20}{\kfieldstrength}$, the orientation at the peak intensity is very similar
for the three tilted angles, 
 which is due to the dominant contribution of oriented adiabatic pendular 
states to the rotational dynamics.

\begin{figure}
 \includegraphics[width=0.95\linewidth]{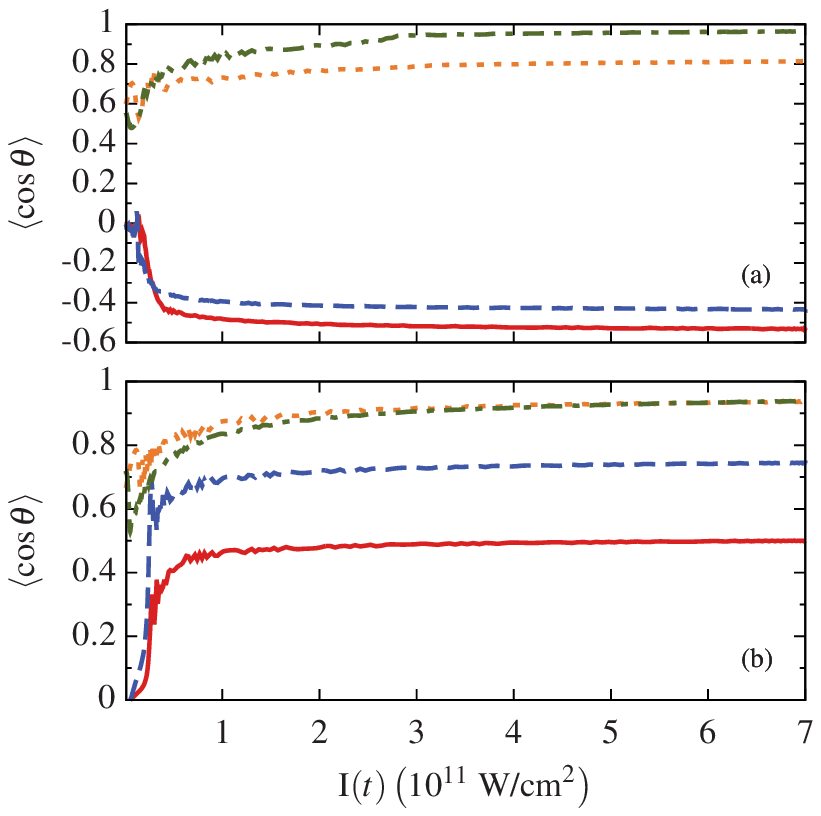}
 \caption{For the $\ptstate{3}{0}{3}{1}$ state, expectation value 
 $\left\langle\cos\theta\right\rangle$  as a function of $\textup{I}(t)$ for a $10$~ns
 laser pulse with peak intensity 
 $\Ialign=\SI{7e11}{\intensity}$ 
for (a)   tilted fields  with $\beta=30\degree$
 and for (b)   parallel fields with the electric field parallel to the LFF $Z$-axis of strength
 $\Estatabs\cos30\degree$.
 The electric field strengths are
  $\Estatabs=\SI{300}{\fieldstrength}$ (red solid line), 
 $\Estatabs=\SI{1}{\kfieldstrength}$ (blue dashed line), 
 $\Estatabs=\SI{10}{\kfieldstrength}$ (orange dotted-line), and
  $\Estatabs=\SI{20}{\kfieldstrength}$  (green dot-dashed line)}
   \label{fig:field_all_parallel}
 \end{figure}

For the $\ptstate{3}{0}{3}{1}$ state, we compare in~\autoref{fig:field_all_parallel} the  orientation 
in tilted fields with the results obtained in a  parallel field configuration which includes only the $Z$-component of the electric field
$\Estatabs\cos\beta$. For this parallel field configuration, the state is oriented, 
$\expected{\cos\theta}$ increases as $\Estatabs$ is enhanced and shows a plateau like 
behavior for $\Inotime(t)\gtrsim\SI{1e11}{\intensity}$.
In contrast, for tilted-fields, the  $\ptstate{3}{0}{3}{1}$ state is anti-oriented for weak dc-fields $\Estatabs=\SI{300}{\fieldstrength}$ and 
$\Estatabs=\SI{1}{\kfieldstrength}$, and oriented for $\Estatabs=\SI{10}{\kfieldstrength}$ and 
$\Estatabs=\SI{20}{\kfieldstrength}$. 
The discrepancy between these results illustrates the importance of the dc-field perpendicular 
to the non-resonant  laser. The non-adiabatic phenomena that take place for tilted fields, 
\ie,  the splitting of the $\ppstate{3}{0}{3}{M}$  manifold and the avoided crossing among 
pendular states having  different field-free magnetic quantum number, strongly affect the 
field-dressed dynamics. For weak electric fields, the impact of these non-adiabatic effects
 is larger and the direction of the orientation is 
reversed. Whereas, for strong dc fields, they show qualitatively similar but quantitative different orientation.

\section{Sources of non-adiabatic effects}
\label{sec:sources}

For tilted fields, the dynamics is characterized by the pendular doublet 
formation, the splitting of the degenerate $\pstate{J}{K_a}{K_c}{M}$-multiplet  at
weak laser intensities, and a large amount of avoided crossings, some of them due to the tilted 
electric field which breaks the azimuthal symmetry.
Since the formation of the pendular doublet has been discussed in detail in
our work on asymmetric top molecules in parallel fields~\cite{omiste:pra_88_2013},
we focus here on exploring the other two non-adiabatic phenomena.

Let us mention that the rotational dynamics of the ground state of each irreducible 
representation is only affected by the pendular doublet formation, as the absolute ground state. 
Thus, for them, it is easy to reach the adiabatic mixed-field orientation 
limit~\cite{omiste:pra_88_2013}. 

\subsection{Coupling in the $J$-manifold}
\label{sub_secion_new}

 \begin{figure*}
 \includegraphics[width=0.95\linewidth]{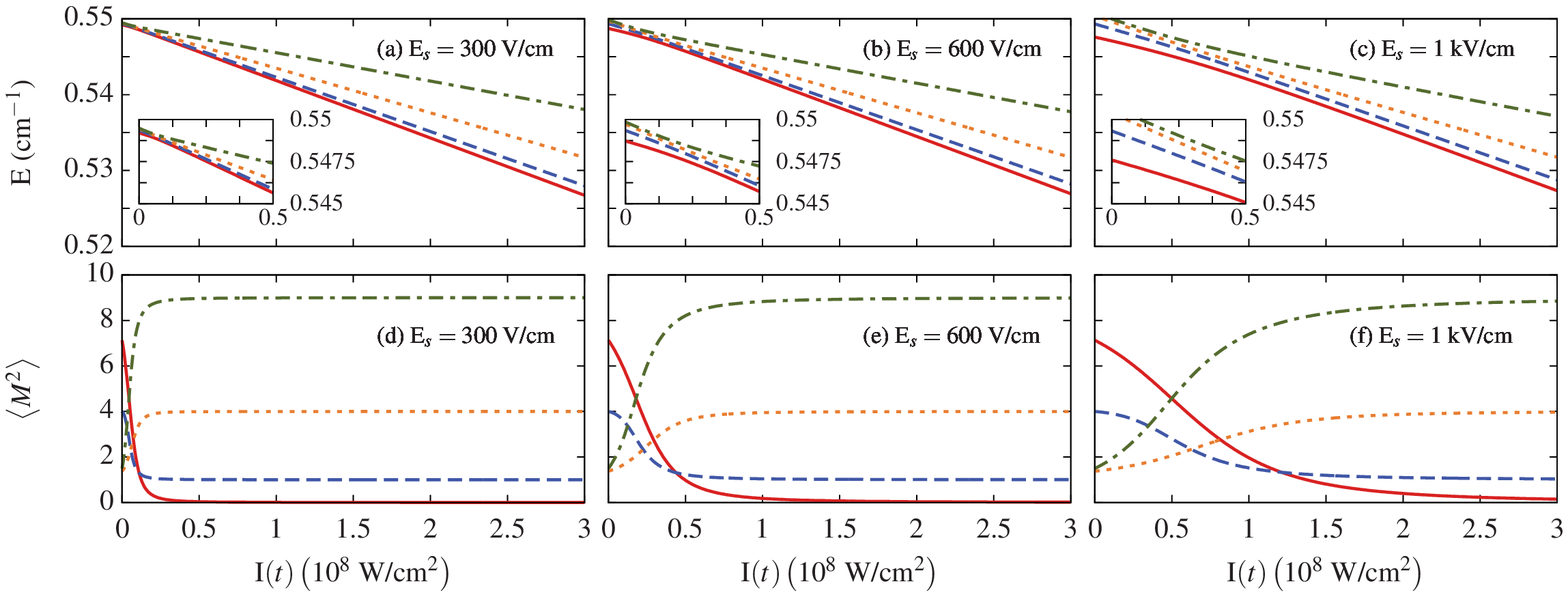}
 \caption{ For the adiabatic pendular states 
 $\ptstate{3}{0}{3}{1}$ (black double-dot-dashed line), 
  $\ppstate{3}{0}{3}{3}$ (red solid line), $\ppstate{3}{0}{3}{2}$ (blue dashed), $\ppstate{3}{0}{3}{1}$ (orange dotted line) and  $\ppstate{3}{0}{3}{0}$ (green dot-dashed),  
  energy and expectation value $\expected{M^2}$
 for a dc field tilted $\beta=30\degree$ and with strength 
$\Estatabs=\SI{300}{\fieldstrength}$~(a) and (d), 
$\Estatabs=\SI{600}{\fieldstrength}$~(b) and (e), 
and 
$\Estatabs=\SI{1000}{\fieldstrength}$~(c) and (f). The laser pulse has $\tau=10$~ns
and peak intensity  $\Ialign=\SI{7e11}{\intensity}$.
}
 \label{fig:multiplet_j_splitting}
 \end{figure*}

In the field-free case, the $\pstate{J}{K_a}{K_c}{M}$
states with $0\le M\le J $ are degenerate due to the azimuthal symmetry. 
The weak electric field of the 
mixed-field orientation experiments breaks their $M$-degeneracy by 
the quadratic Stark splitting, $\Delta E \sim \Estatabs^2$.
For $\Estatabs=\SI{300}{\fieldstrength}$ and $\beta=30\degree$, 
the neighboring levels of the  $\pstate{3}{0}{3}{M}$-manifold are separated by
$1.6\times 10^{-4}$~cm$^{-1}$, $9.4\times 10^{-5}$~cm$^{-1}$
 and $3.1\times 10^{-5}$~cm$^{-1}$.
Due to these small energy gaps, even a weak laser field provokes strong couplings among 
them, which significantly affect the rotational dynamics.  
We consider a  dc field tilted $\beta=30\degree$ with strengths 
$\Estatabs=\SI{300}{\fieldstrength}$, $\SI{600}{\fieldstrength}$, and $\SI{1000}{\fieldstrength}$. 
In \autoref{fig:multiplet_j_splitting} we present the energy, and expectation value
 $\expected{M^2}$ of the adiabatic pendular states $\ppstate{3}{0}{3}{M}$ 
as the laser intensity is increased till $\Inotime(t)=\SI{3e8}{\intensity}$.
In the presence of only an electric field forming an angle $\beta$ with  the LFF $Z$-axis, 
the projection of $\mathbf{{J}}$ along the dc-field axis is a good quantum number,
 but not along the LFF $Z$-axis, cf.~\autoref{fig:multiplet_j_splitting}~(d), (e), and (f). 
As the laser intensity increases, the levels suffer several avoided crossings,
whose widths are larger for stronger dc-fields, see~\autoref{fig:multiplet_j_splitting}~(a)-(c).
The effects of these avoided crossings are recognized  in the time  evolution of $\expected{M^2}$.
After them, $\expected{M^2}$ shows a constant behavior as $\Inotime(t)$ is increased. 
Indeed, when the interaction due to the laser field is dominant, the 
projection  of $\mathbf{J}$ along the LFF $Z$-axis becomes a quasi good quantum number,
as is observed in~\autoref{fig:multiplet_j_splitting}~(d), (e) and (f).

 \begin{figure}
 \includegraphics[width=0.95\linewidth]{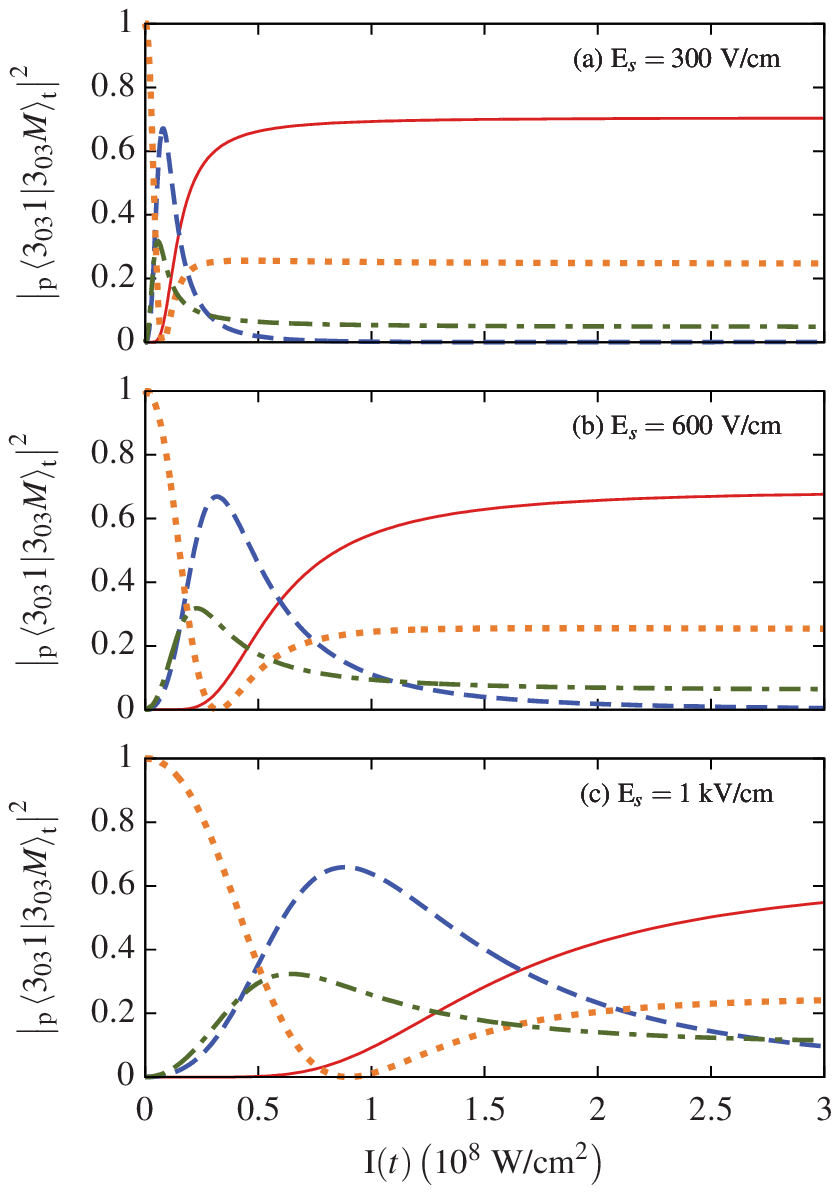}
 \caption{The squares of the projection of the time-dependent wave function $\ptstate{3}{0}{3}{1}$ onto the adiabatic states of the $\pstate{3}{0}{3}{M}$  multiplet, 
 \ie, $\ppstate{3}{0}{3}{3}$, $\ppstate{3}{0}{3}{2}$, $\ppstate{3}{0}{3}{1}$ and  $\ppstate{3}{0}{3}{0}$ as a function of $\Inotime(t)$. The field configurations are
  $\Ialign=\SI{7e11}{\intensity}$, $\tau=10$~ns, $\beta=30\degree$ and 
  (a)  $\Estatabs=\SI{300}{\fieldstrength}$, 
  (b)  $\Estatabs=\SI{600}{\fieldstrength}$, and 
  (c)  $\Estatabs=\SI{1000}{\fieldstrength}$. 
  The labels of the states are the same as in~\autoref{fig:multiplet_j_splitting}.}
 \label{fig:multiplet_j_splitting_es}
 \end{figure}
The couplings within the states of the $\ppstate{J}{K_a}{K_c}{M}$-manifold  have a strong impact on the  
rotational dynamics since,
for a given state,  the time-dependent  wave function might have  contributions 
 from all the adiabatic pendular   states within this multiplet. 
This is illustrated  in~\autoref{fig:multiplet_j_splitting_es}
 with the weights of the adiabatic pendular states  $\ppstate{3}{0}{3}{M}$
into the time-dependent wave function of  $\ptstate{3}{0}{3}{1}$
for $\Estatabs=\SI{300}{\fieldstrength}$, $\SI{600}{\fieldstrength}$ and
$\SI{1000}{\fieldstrength}$, $\beta=30\degree$
and a $10$~ns laser pulse with   peak intensity $\Ialign=\SI{7e11}{\intensity}$.
As $\Inotime(t)$ increases,  the weights of the adiabatic pendular states change drastically and  are 
redistributed within the  manifold.
This population transfer depends on the coupling among the states in the multiplet,  
on the initial energy gap between them, which is determined by 
the dc-field strength and the angle $\beta$, and on the way the alignment pulse is turned on.
The field-dressed dynamics in this region is characterized by large time scales and very large 
adiabaticity ratios $\eta\gg 1$ among the adiabatic pendular states~\cite{omiste:pra2012}. 
As a consequence, very long laser pulses are required to reach  the adiabatic limit. 
Once the manifold is split, the weights reach a plateau-like behavior, which is kept till one 
of the adiabatic pendular states suffers an avoided crossing, which  might occur at stronger 
laser intensities.

\subsection{Check of the diabatic model}
\label{subsection_aaa}
In this section, we check the validity of the diabatic model, which is based
 on the  weak coupling, induced by the  tilted weak electric field, between states with different field-free 
magnetic quantum numbers~\cite{omiste:pccp2011}.
In the presence of only a laser field, the magnetic quantum number of the field-dressed states is conserved,
 and by adding a  tilted static electric field 
 the rotational symmetry around the laser polarization axis of the ac-field Hamiltonian is broken.
For a weak dc field, the  interaction due to this dc field could be considered as a perturbation to the ac-field Hamiltonian,
and, in this case, $\left\langle M^2\right\rangle$ is almost conserved. 
The diabatic model  assumes that an avoided crossing is crossed adiabatically
(diabatically) if  the  two involved  states have the same (different) $\expected{M^2}$.
Then, the adiabatic model implies that  $\expected{M^2}$ should approximately remain constant.  

 \begin{figure}
 \includegraphics[width=0.95\linewidth]{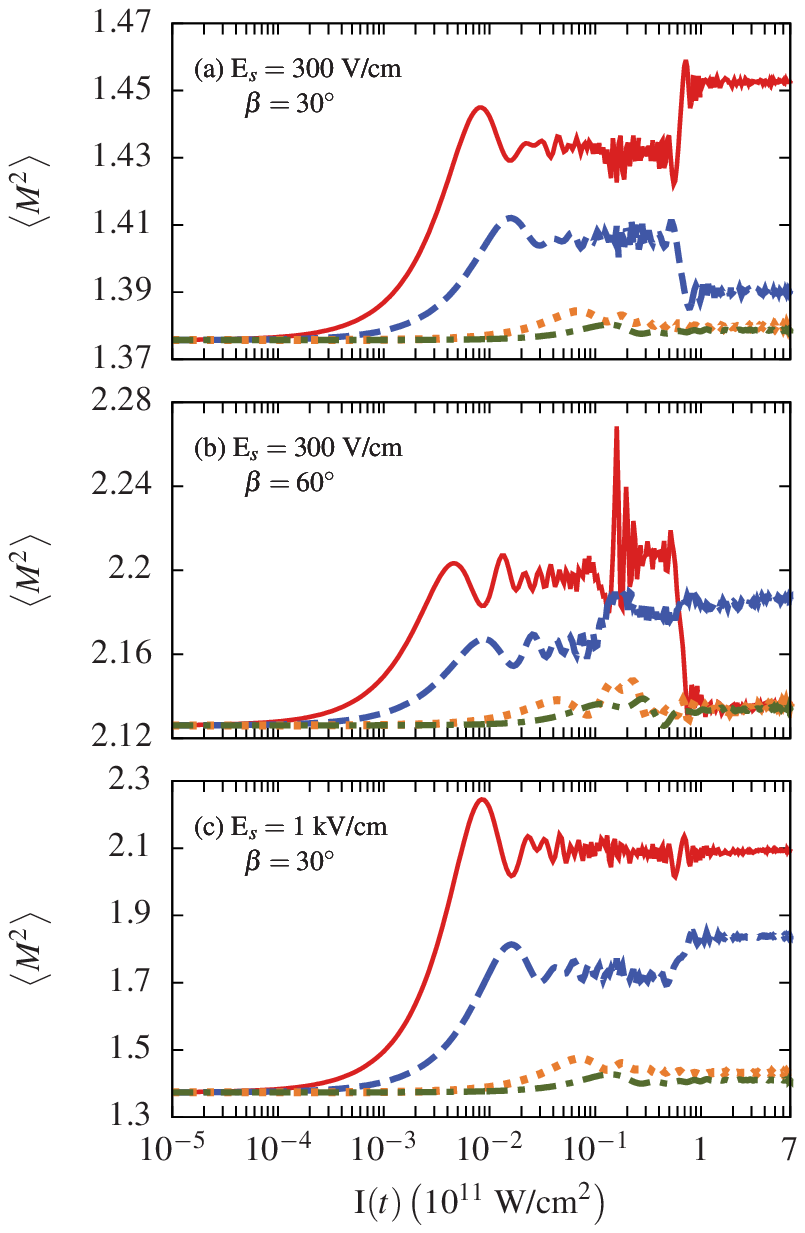}
 \caption{For the time-dependent state $\ptstate{3}{0}{3}{1}$,   
expectation $\expected{M^2}$  as a function of $\Inotime(t)$
 for the rotational state 
 for  (a) $\Estatabs=\SI{300}{\fieldstrength}$ and 
 $\beta=30\degree$ (b) 
 $\Estatabs=\SI{300}{\fieldstrength}$ and $\beta=60\degree$ and (c) $\Estatabs=\SI{1000}{\fieldstrength}$ and 
 $\beta=30\degree$. The laser pulse has $\Ialign=\SI{7e11}{\intensity}$,  
 and  temporal widths $\tau=10$~ns 
 (red solid line), $5$~ns (blue dashed line), $1$~ns (orange dotted line) and $0.5$~ns (green dot-dashed line).
 \label{fig:m2_time}}
 \end{figure}
As an example,  we present in~\autoref{fig:m2_time} the time evolution of  $\expected{M^2}$ 
for the  $\ptstate{3}{0}{3}{1}$  state in several field configurations. 
Before the pulse is turned on, the initial state is the adiabatic pendular level  of the 
corresponding  dc-field configuration,
and, as indicated above,  $\expected{M^2}$ differs from 
its  field-free value due to the tilted electric field.
As $\Inotime(t)$ increases, $\expected{M^2}$  increases and reaches a maximum,
which appears at lower intensities for longer pulses.  
If the pulse is  short, the rotational  dynamics does not  adapt to the
 time-dependent interaction, which provokes non-adiabatic avoided
crossings, and $\expected{M^2}$ does not change significantly.
For longer pulses, the field-dressed dynamics is more adiabatic and  larger changes on 
$\expected{M^2}$ are observed at moderate laser intensities, and, therefore, $\expected{M^2}$ reaches a  larger value at the maximum.
By increasing the dc-field strength, the Stark  couplings among neighboring levels are larger provoking  
larger changes in $\expected{M^2}$, see~\autoref{fig:m2_time}(a) and (c). 
After the first maximum, $\expected{M^2}$ shows a  rapid oscillatory behavior
due to the presence of several avoided crossings, which are crossed diabaticaly
transferring part of the population. 
For $\tau=10$~ns, $\Estatabs=\SI{300}{\fieldstrength}$, $\beta=30\degree$ and $\beta=60\degree$,
$\expected{M^2}$ shows a sudden change 
due to highly non-adiabatic avoided crossing among states with different values of $\expected{M^2}$.
For $\Inotime(t)\gtrsim\SI{e11}{\intensity}$,  $\expected{M^2}$  reaches a 
 plateau-like behavior with small fluctuations, and, therefore, in this region the diabatic model provides a good approximation
 to the field-dressed dynamics.

These results show that the failure of  the diabatic model mainly occurs at low or moderate laser field
intensities. In this regime, the field dressed states are strongly coupled and several non-adiabatic effects take place
resulting on a time-dependent  wave function which is a linear combination of the eigenstates of the intantaneous field-dressed
Hamiltonian. These non-adiabatic effects cannot be captured by the 
diabatic  model, since it considers an unambiguous correspondence between  eigenstates.

\section{Conclusions}
\label{sec:conclusions}

In this work, we have investigated the rotational dynamics of asymmetric-top molecules  on  a
tilted-field  configuration similar to those used in current mixed-field orientation experiments.
By considering the benzonitrile molecule as prototype, the richness and variety
of the field-dressed dynamics have been illustrated. 
We have addressed unique non-adiabatic effects  of the tilted field configuration such
as the $J$-multiplet splitting and the coupling between states with different field-free magnetic 
quantum numbers.  
By increasing the dc-field strength, the energy spacings among the states on a $J$-manifold and
on the quasidegenerate pendular doublets are enhanced. Thus, the characteristic time scales of these two non-adiabatic phenomena are 
reduced easing the experimental requirements for an adiabatic dynamics. 
However, the large amount of narrow avoided crossings that  emerge for moderate and strong laser intensities frustrates
 the hunt of an adiabatic field-dressed dynamics for rotationally excited states. 
As a consequence, for excited rotational states, it becomes more challenging to experimentally 
reach the  adiabatic limit.

In this time-dependent framework, we have revisited  the mixed-field orientation experiment  
of a state-selected molecular beam of benzonitrile~\cite{PhysRevA.83.023406}. 
Our analysis includes  $92.5\%$ of  the molecular beam with the experimental 
weights, and the experimental field configuration: 
a weak static electric field combined with a non-resonant linearly polarized laser pulses. 
In this time-dependent description, the  degree of orientation of the molecular ensemble is 
smaller than the experimentally measured~\cite{PhysRevA.83.023406} and similar to the orientation provided by the 
diabatic model~\cite{omiste:pccp2011}.
By completing the molecular beam with the rest of populated states in the experiment and  taking into
account the volume effect, this time-dependent  orientation  ratio should not be significantly 
modified and should not become closer to the experimental one.
The disagreement between the theoretical and  experimental results could be due to the Coulomb explosion, and the
subsequent detection of the molecular 
ions, and the way these processes are simulated or to the lack of adiabaticity on previous steps 
of the experiment, such as the state selection, which might modify the experimental weights of the
rotational states in the molecular beam.

A rather natural extension of this work would be to investigate the rotational dynamics 
of an  asymmetric-top molecule without rotational symmetry. 
This time-dependent study should allow us to review  the conclusions of the 
adiabatic analysis of the 6-chloropyridazine-3-carbonitrile (CPC) in combined electric and
non-resonant laser fields~\cite{Hansen2013}.

\begin{acknowledgments}
We would like to thank 
Jochen K\"upper,  Henrik Stapelfeldt, and Linda V. Thesing for fruitful discussions, and 
Hans-Dieter Meyer for providing us the code of the short iterative Lanczos algorithm. 
Financial support by the Spanish project FIS2014-54497-P (MINECO), and the Andalusian 
research group FQM-207 is gratefully appreciated. JJO acknowledges the support by the Villum Kann Rasmussen (VKR) Center of Excellence QUSCOPE.

\end{acknowledgments}
\bibliographystyle{apsrev4-1}
\bibliography{asymmetric_time}

\end{document}